\documentclass[1p]{elsarticle}
\usepackage{amssymb}

\def\ii{\'{\i }}
\usepackage{amssymb}
\begin{document}
\begin{frontmatter}
\title{Long range correlations, event simulation and parton percolation}
\author{C. Pajares}
\address{IGFAE and Departamento
de F\ii sica de Part\ii culas, Univ. of Santiago de Compostela,
15706, Santiago de Compostela, Spain}
\begin{abstract}
We study the RHIC data on long range rapidity correlations,
comparing their main trends with different string model
simulations. Particular attention is paid to color percolation
model and its similarities with color glass condensate. As both
approaches corresponds, at high density, to a similar physical
picture, both of them give rise to a similar behavior on the
energy and the centrality of the main observables. Color
percolation explains the transition from low density to high
density.
\end{abstract}
\end{frontmatter}
\section{Introduction}
Forward- backward multiplicity correlations have been studied
recently in RHIC experiments
\cite{:2009dqa}\cite{Back:2006id}\cite{Adler:2007fj} STAR data
\cite{:2009dqa} has measured the parameter b, defined by
\begin{equation}
b=\frac{<n_{F}n_{B}>-<n_{F}><m_{B}>}{<n_{F}^{2}>-<n_{F}>^{2}}=\frac{D_{BF}^{2}}{D_{FF}^{2}}
\end{equation}
where $D_{BF}$ and $D_{FF}$ are the backward-forward and
forward-forward dispersions respectively. In between the forward
and backward rapidities bins there is a rapidity gap $\Delta
\eta$, being measured b as a function of $\Delta \eta$. FB
multiplicity correlations probe the longitudinal characteristics
of the system produced in AA and hh collisions. The presence of
sizable long range rapidity correlations was predicted in some
particle models, as the Dual Parton Model (DPM)
\cite{Capella:1992yb}\cite{Amelin:2001sk} and the Color Glass
Condensate (CGC) model
\cite{Armesto:2006bv}\cite{Kovchegov:1999ep}\cite{Dusling:2009ni}
that contain longitudinal color flux tubes in CGC. The STAR data
at 200 GeV show that from most central $(0-10\%)$ to mid-
peripheral $(40-50\%)$ Au-Au collisions there is a general
decrease in the FB correlation strength b. The shape of b for
$(40-50\%)$ centrality as a function of $\Delta \eta$ indicates a
fast decreasing with increasing $\Delta \eta$. If only short range
correlations contributed to b, all centralities should resemble
the $(40-50\%)$ result, however b remains flat in the whole range
of $\Delta \eta$ even for $\Delta\eta>1.5$ for $20-30\%$,
$10-20\%$, $0-10\%$ centralities, indicating long range
correlation. In pp the b value is much smaller, decreasing with
increasing $\Delta \eta$, being close to zero for $\Delta
\eta>1.2$ .

 Not all the strings models
can reproduce data. First of all the strings must be extended in
both hemispheres. There are models like Hijing, that the strings
are formed between the partons of the same nucleon, there is not
color exchange between projectile and target, and the fragmentation of
these strings gives rise to particles that most of them are produced
only in one rapidity hemisphere and therefore it is not able to
reproduce the b shape.
The inclusion of rescattering between partons of different
nucleons, as it is done in PACIAE model (based on HIJING) improves the comparison
with the experimental data and it can
reproduce the strength and shape of b for central but not for
peripherical collisions \cite{Yan:2009zzd}.

In modified wounded nucleon of Bzdak \cite{Bzdak:2009dr} is
explained the enhancement of b with respect to the elementary pp
interaction by the asymmetric shape of the rapidity density of
produced particles from a single wounded nucleon in addition to
the fluctuations in the number of wounded nucleons in the
colliding nuclei. This asymmetric shape means that the particles
are not produced only in one hemisphere. However, in this model
the length of the correlations is the same for pp than for AA
central collisions, contrary to the experimental data. The string
percolation \cite{Armesto:1996kt}\cite{Braun:1999hv} and the CGC
models \cite{McLerran:1993ni}\cite{Lappi:2006fp} have the right
longitudinal structure able to explain at least qualitatively, the
dependence on the energy and centrality of the strength and length
of the long range rapidity correlations. Both of them predict a
very long rapidity correlations for AA collisions at LHC energies,
indeed more than eight units of rapidity. As far as the length of
the long range correlations establishes the length of the
near-side ridge structure, it is predicted a very long ridge at
LHC.

\section{The string percolation model}
Multiparticle production is currently described in terms of color
strings stretched between the partons of the projectile and the
target. These strings decay into new ones, via $q-\bar{q}$
production, (Schwinger mechanism), and subsequently hadronize to
produce observed hadrons. Color strings may be viewed as small
areas in the transverse space, $S_{1}= \pi r_{0}^{2}$, with $r_{0}
\simeq .2-.3$ fm, filled with the color field created by the
colliding partons. With increasing energy and/or atomic number of
colliding particles, the number of exchange string grows, and they
start to overlap, forming clusters, very much like disk in 2-
dimensional percolation. Each cluster has a higher color field and
an energy- momentum that correspond to the sum of the energy
momentum of the overlapping strings. At a certain critical density
$\eta_{c}\simeq
 1.2-1.5$ a macroscopical cluster appears which marks the
percolation phase transition. The variable
$\eta$ used above is defined by $\eta_{c} = N_{s}\frac{ S_{1}}{ S_{A}}$,
where
$N_{s}$ is the number of strings, which is proportional to the number of collisions
$N_{s} \sim N_{A}^{\frac{4}{3}}$, and
$S_{A}$
is the overlapping area. At $b=0$, $S_{A}=\pi R_{A}^{2}$.

 The color field $\overrightarrow{Q}_{n}$ inside the cluster is the
vectorial sum of the color field $\overrightarrow{Q_{1}}$ of each
individual string. The resulting color field covers the area
$S_{n}$ of the cluster. As
$\overrightarrow{Q_{n}^{2}}=(\sum\overrightarrow{Q_{1}})^{2}$, and
the individual string colors may be oriented in an arbitrary
manner respective to one another, the average
$\overrightarrow{Q_{1i}}\overrightarrow{Q_{1j}}$ is zero and
$\overrightarrow{Q_{n}^{2}}=n\overrightarrow{Q_{1}}$
$\overrightarrow{Q_{1}}$ depends also on the area $S_{1}$ as well
on the total area of the cluster $ S_{n}$, therefore
\begin{equation}
Q_{n}=\sqrt{\frac{nS_{n}}{S_{1}}}Q_{1}
\end{equation}
in such a way that if the strings are just touching each other
$S_{n}=nS_{1}$ and $Q_{n}=nS_{1}$, so the strings are independent of each
other. On the contrary if they are fully overlap, $S_{n}=S_{1}$ and
$Q_{n}=\sqrt{n}Q_{1}$. from (2) we can deduce that the mean multiplicity
$\mu_{n}$ and the mean $<p_{T}^{2}>_{n}$ of the particles produced by the
cluster of n strings.
\begin{equation}
\mu_{n}=\sqrt{\frac{nS_{n}}{S_{1}}} \mu_{1}, \mbox{   }
<p_{T}^{2}>_{n}=\sqrt{\frac{nS_{1}}{S_{n}}} <p_{T}^{2}>_{1}
\end{equation}
As we have obtained \cite{Braun:1999hv}.
\begin{equation}
<n\frac{S_{1}}{S_{n}}>=\frac{\eta}{1-e^{-\eta}}=\frac{1}{F(\eta)^{2}}
\end{equation}
where $1-e^{-\eta}$ is the fraction of the total area covered by
the strings, we can write for the multiplicity and $<p_{T}^{2}>$
\begin{equation}
\mu_{n}=N_{s}F(\eta)\mu_{1},\mbox{  } <p_{T}^{2}>=\frac{<p_{T}^{2}>_{1}}{F(\eta)}
\end{equation}
At high density, $F(\eta)\simeq {A}^{-1/3}$ and $\mu_{n}\simeq
 N_{A}$, i. e. it is obtained the saturation of the multiplicity per participant.
If $N_{s}\simeq s^{2\lambda}$, $\mu_{n}\simeq s^{\lambda}$, i.e.
grows with the energy slower than $N_{s}$. Outside the
midrapidity, $N_{s}$ is proportional to $N_{A}$ instead of
$N_{A}^{4/3}$. Therefore there is an additional suppression factor
$N_{A}^{1/3}$ compared to central rapidity.

 The second
of equation (5) establishes the transverse size correlations which is proportional
to $\frac{F(\eta)}{<p_{T}^{2}>_{1}}$ or  $r_{0}^{2}F(\eta)$.

\section{String percolation and Color Glass Condensate.}

The string percolation and the glasma formed from the Color Glass
Condensate have many similarities giving rise to similar
predictions, as we are going to show.

One of the main scales of
both models is the transverse size given by $r_{0}^{2}F(\eta)$ and
$\frac{1}{Q_{s}^{2}}$ respectively which in the high density and
high energy limit, both behave as $N_{A}^{-1/3}$ and
$s^{-\lambda}$.

The effective number of independent color sources, $<N>$ is
given by the ratio between the surface occupied by strings and
the transverse size
\begin{equation}
<N>=\frac{(1-e^{-\eta})R_{A}^{2}}{r_{0}^{2}F(\eta)}=(1-e^{-\eta})^{\frac{1}{2}}\eta^{\frac{1}{2}}(\frac{R_{A}}{r_{0}})^{2}
\end{equation}
which behaves at low density, (energy) as
\begin{equation}
\eta(\frac{R_{A}}{r_{0}})^{2}, \mbox{   i.e.     }
N_{A}^{\frac{4}{3}} , \sqrt{s}^{2\lambda}
\end{equation}
and at high density, (energy)
\begin{equation}
\sqrt{\eta}(\frac{R_{A}}{r_{0}})^{2}, \mbox{  i. e.     } N_{A},
\sqrt{s}^{\lambda}
\end{equation}
In the Glasma we have $k=Q_{s}^{2}R_{A}^{2}$ color flux tubes each
producing $\frac{1}{\alpha_{s}}$ gluons in such a way that the
number of color flux tubes k, behaves as $N_{A}$, $s^{\lambda}$ in
agreement with (8) \cite{Kharzeev:2004if}\cite{DiasdeDeus:2005}.
 The origin of this behaviour in CGC, that gives rise
to saturation is the non-linear behaviour of the strong color fields at
low x produced by the color sources (gluons) located at high x, in
percolation, this behaviour is also due to the non-linear superposition of
the strength of the color fields, which gives rise to a factor $n^{1/2}$
instead of n in the equation (2).

The role of the running coupling constant
$\frac{1}{\alpha_{s}}$ of CGC is played in percolation of strings
by the factor $\sqrt{1-e^{-\eta}}$, as it can been seen from the
comparison of multiplicities in both models, i. e.
$\frac{1}{\alpha_{s}}Q_{s}^{2}R_{A}^{2}$
 and equation (5). As the density, or energy, increases both
$\frac{1}{\alpha_{s}}$ and $(1-e^{-\eta})^{1/2}$ also increase.
The exponent $\lambda$ in formula (7) and (8) has been determined,
using energy momentum conservation, obtaining
$\lambda=\frac{2}{7}$ \cite{DiasdeDeus:2005} value close to the
one obtained in \cite{Albacete:2004gw}, using CGC.

The rapidity extension of the color flux tubes of the glasma are
$\frac{1}{\alpha_{s}}$ which grows at high density as $ln(N_{A})$,
$ln(s)$. In percolation, as the energy momentum of the clusters is
the sum of the individual energy momentum of the strings, the
rapidity extension of a cluster formed by $N_{s}$ strings is $\Delta y_{N}=\Delta
y_{1}+2ln(N_{s})$,
which grows also as $ln(N_{A})$, $ln(s)$. Notice, that due to that both
CGC and
percolation predict a very long rapidity correlation al LHC, b will be flat up to
$\Delta \eta \geq 8$.

The inverse of the normalized fluctuations as the number of
effective color sources
\begin{equation}
\bar{K}\equiv\frac{<N>^{2}}{<N^{2}>-<N>^{2}}
\end{equation}
is infinity at low and high density, having a minimum at one
intermediate value of $\eta$. In fact
\begin{equation}
\bar{K}\equiv\frac{<N>}{(1-e^{-\eta})^{3/2}}
\end{equation}
that behaves at low density as
$\eta^{-\frac{1}{2}}(\frac{R_{A}}{r_{0}})^{2}$ and at high
density as $\eta^{\frac{1}{2}}(\frac{R_{A}}{r_{0}})^{2}$ that
grows like $N_{A}$ and $s^{\lambda}$. The number k of color flux
tubes of the glasma we have seen already that has similar
behavior.

In both models, the multiplicity distributions are described by
negative binomial distributions \cite{Gelis:2009wh}
\cite{DiasdeDeus:2003ei}. This distributions are determined by two
parameters: the mean multiplicity and the inverse of the
normalized fluctuations of the multiplicities, i. e
$<n>^{2}/(<n^{2}>-<n>^{2})$. In the glasma this parameter is
$\bar{k}=<N>k_{0}$, where $k_{0}$ is the corresponding parameter
of the negative binomial distribution for one string, At low
density,$\bar{k}\rightarrow \infty$. The value $k_{0}=\infty$ means
that for one string the negative binomial is a Bose-Einstein
distribution. This result coincides with what is found in the
Glasma.

Therefore k, at low density decreases with density (energy) but as
for the density increases, k increases with the density ( energy).
This behavior means than the width of the distribution $<n>P_{n}$
as a function of $\frac{n}{<n>}$ should start to become narrower
from one determined density (energy). In AA collisions at RHIC, we
are in the high density sector and in fact k increases with the
centrality.

In pp collisions, in the framework of string percolation, we
expect a change on the behavior of k with the energy as far as we
go from low density to high density. The scale can be established
from the AA case as follows.
\begin{equation}
\eta_{AA}=N_{s}\frac{r_{0}^{2}}{R_{A}^{2}}=\frac{N_{A}^{\frac{4}{3}}}{N_{A}^{\frac{2}{3}}}(\frac{r_{0}}{R_{p}})^{2}N_{p}^{s}
\end{equation}
where $N_{s}$ and $N_{p}^{s}$ are the number of strings exchange in AA and pp
collisions respectively. Therefore
\begin{equation}
\eta_{AA}^{s}=N_{A}^{\frac{2}{3}}\eta_{pp}^{s}
\end{equation}
assuming that the critical value $\eta_{c}=1.2-1.5$ is reached in
Pb-Pb central collisions at an energy between SPS and RHIC, and
taking into account the $s^{2\lambda}$ dependence of $N_{p}^{s}$,
we obtain that in pp the critical value is reached for an energy
in between 6-14 TeV. Experimentally, in pp collisions k is
decreasing from SPS up to 2.36 TeV. We expect a change on this
behavior at an energy close to 7 TeV \cite{DiasdeDeus:2004qk}.

The forward-backward correlation b, that can be written as
\begin{equation}
b=\frac{1}{1+\frac{k}{<n>_{F}}}
\end{equation}
In percolation
\cite{Brogueira:2009nj}\cite{Brogueira:2006yk}\cite{Amelin:1994mf}
from eqs. (10) and (5) we obtain
\begin{equation}
b=\frac{1}{1+\frac{d}{(1-e^{-\eta})^{3/2}}}
\end{equation}
which at low density vanishes, and at high density grows becoming $\frac{1}{1+d}$ being d a constant independent on
the density and energy.

In CGC b is given by \cite{Armesto:2006bv}\cite{Kovchegov:1999ep}
\begin{equation}
b=\frac{1}{1+\alpha_{s}^{2}c}
\end{equation}
that also grows with density and$/$or energy.

The strength and length of b are related to the high and length of the
ridge structure of near side correlations. The percolation and the
CGC prediction of long range correlations with length larger than eight
units of rapidity means that the ridge structure will be extremely long at
LHC energies.

Finally, let us mention that from the transverse size
correlations and the number of effective color sources we deduce the shear
viscosity which is given by
$<p_{T}>/nS_{1}$ \cite{DiasdeDeus:2006xk}
 where n is the
number of effective color sources per volume, i. e.
\begin{equation}
n=\frac{<N>}{\pi R_{A}^{2} L}
\end{equation}
where L is the longitudinal extension, $L\simeq 1$ fm. From (3) and (6) we
obtain that the shear viscosity is
\begin{equation}
s.v =<p_{T}>_{1}L\eta^{\frac{1}{4}}\frac{1}{(1-e^{-\eta})^{\frac{5}{4}}}
\end{equation}
which behaves like $\eta^{-1}$ at low $\eta$ and grows like
$\eta^{\frac{1}{4}}$ at high density, having a minimum close to
the critical value of percolation.

Recently, in string percolation has been computed the equation of state.
The results for the energy density, speed sound, energy density and
pressure \cite{Scharenberg} are just on the same curve computed in
lattice QCD.

All the similarities seen on
the behavior of the main observables with the density and energy
in string percolation and CGC, including the prediction of very
long range rapidity correlations, indicate that both models are
complementary views of the same physical picture.

\section{Acknowledgements}
Most of the work reported here was done in collaboration with J. Dias de Deus.
I thank also J. G. Milhano and I. Bautista for helping me in this work.
 I thank the support of the project FPA2008-01177 of Spain, and of
the Xunta de Galicia.

\end{document}